\documentclass[11pt]{article}
\usepackage{amssymb}
\usepackage{amsmath}

\usepackage{acl}

\usepackage{times}
\usepackage{latexsym}
\usepackage{booktabs} 
\usepackage{multirow} 

\usepackage[T1]{fontenc}

\usepackage[utf8]{inputenc}

\usepackage{microtype}

\usepackage{inconsolata}

\usepackage{graphicx}

\title{Audio ControlNet for Fine-Grained Audio Generation and Editing}

\author{
 \textbf{Haina Zhu\textsuperscript{1,2}},
 \textbf{Yao Xiao\textsuperscript{1}},
 \textbf{Xiquan Li\textsuperscript{1}},
 \textbf{Ziyang Ma\textsuperscript{1}},
\\
 \textbf{Jianwei Yu\textsuperscript{4}},
 \textbf{Bowen Zhang\textsuperscript{3}},
 \textbf{Mingqi Yang\textsuperscript{3}},
 \textbf{Xie Chen\textsuperscript{1,2}}\footnote{Corresponding author.},
\\
 \textsuperscript{1}X-LANCE Lab, School of Computer Science, Shanghai Jiao Tong University, \\
 \textsuperscript{2}Shanghai Innovation Institute,
 \textsuperscript{3}MiniMax,
 \textsuperscript{4}Independent Researcher
\\
 \{hainazhu, chenxie95\}@sjtu.edu.cn
 \\
 \\
 \textbf{Code:}  \url{https://github.com/juhayna-zh/AudioControlNet}
 \\
 \textbf{Demo:}   \url{https://audio-controlnet.github.io}
}

\begin{document}
\maketitle
\begingroup
\renewcommand{\thefootnote}{\fnsymbol{footnote}}
\setcounter{footnote}{0}
\footnotetext{* Corresponding author.}
\endgroup

\begin{abstract}
We study the fine-grained text-to-audio (T2A) generation task. While recent models can synthesize high-quality audio from text descriptions, they often lack precise control over attributes such as loudness, pitch, and sound events.
Unlike prior approaches that retrain models for specific control types, we propose to train ControlNet models on top of pre-trained T2A backbones to achieve controllable generation over loudness, pitch, and event roll.
We introduce two designs, T2A-ControlNet and T2A-Adapter, and show that the T2A-Adapter model offers a more efficient structure with strong control ability. With only 38M additional parameters, T2A-Adapter achieves state-of-the-art performance on the AudioSet-Strong in both event-level and segment-level F1 scores.
We further extend this framework to audio editing, proposing T2A-Editor for removing and inserting audio events at time locations specified by instructions.
Models, code, dataset pipelines, and benchmarks will be released to support future research on controllable audio generation and editing.
\end{abstract}

\section{Introduction}

Text-to-audio (T2A) generation aims to synthesize general audio and sound effects from natural language descriptions, which has recently attracted increasing attention due to its potential applications in sound design \cite{evans2025sao}, music creation \cite{copet2023musicgen}, and video production \cite{cheng2025mmaudio}. With the rapid development of generative modeling techniques, particularly diffusion-based models, recent T2A systems \cite{evans2025sao, liu2023audioldm, liu2024audioldm2, hai2024ezaudio} have demonstrated remarkable progress in both audio quality and text-audio alignment. 

\begin{figure}[t]
  \centering
  \includegraphics[width=0.95\linewidth]{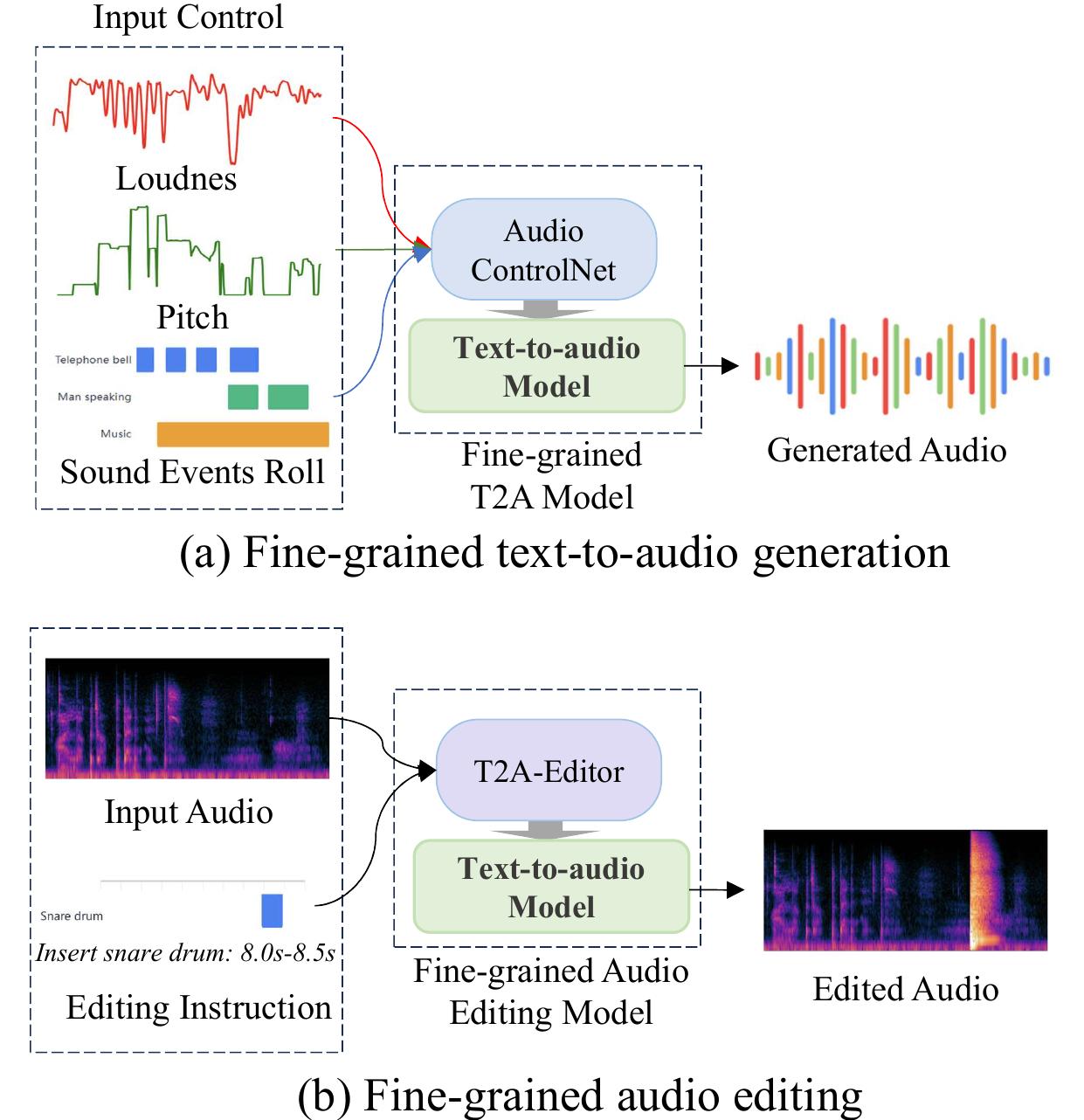}
  \caption{An illustration of fine-grained text-to-audio (T2A) generation tasks and fine-grained audio editing tasks, along with the proposed Audio ControlNet.}
  \label{fig:tasks}
\end{figure}

However, despite these advances, most T2A models offer only coarse-grained control via text and struggle to enforce precise temporal event structures, such as specifying that a sound (e.g., a dog bark) should occur between 2 and 4 seconds, or that multiple sound events should be active simultaneously, as illustrated in Figure \ref{fig:tasks}. They also provide limited signal-level control, such as loudness dynamics, pitch contours, or specific pitch components.

More recently, several approaches have attempted to achieve finer-grained control in T2A generation. These methods generally fall into two categories. The first uses textual descriptions to specify when and which sounds should occur \cite{wang2025audiocomposer, jiang2025controlaudio}. While intuitive, this places high demands on the model’s language understanding, and misinterpreted timings can lead to imprecise control. The second employs structured embeddings to explicitly represent event occurrences \cite{zheng2025picoaudio2, xie2024picoaudio}, allowing more precise temporal control but often requiring training from scratch and complex data simulation pipelines.

It is also worth noting that most approaches remain limited in their ability to extend control to finer signal-level attributes, such as precise loudness and pitch variations. This motivates the need for a framework that can achieve precise, fine-grained control while reducing training complexity and preserving the flexibility of pre-trained models.

Rather than retraining entire models, we focus on lightweight auxiliary networks that enable existing T2A models to accept fine-grained control signals, preserving generative capacity while reducing training complexity. Inspired by ControlNet in text-to-image generation \cite{zhang2023adding, ye2023ip, mou2024t2i, zhao2023unicontrolnet}, we introduce \textbf{Audio ControlNet }models for T2A models, exploring two complementary designs: a copy-network-based architecture, \textbf{T2A-ControlNet}, and a lightweight-encoder-based architecture, \textbf{T2A-Adapter}.

We further argue that fine-grained control signals should be structured rather than text-based, and this requires a representation that is both flexible and extensible to accommodate diverse control conditions. To this end, we formulate all control inputs as temporal sequences aligned with the audio timeline. Within this unified framework, we design \textbf{structured representations and feature extractors} for loudness, pitch, and sound events.

We further extend our framework to fine-grained audio editing, where instructions specify both content and exact time spans (e.g., “insert clapping from 3.0s to 4.0s,” “remove speaking from 6.0s to 8.0s”). Following the same auxiliary-network paradigm, we propose \textbf{T2A-Editor}, which equips existing T2A models with precise insertion and removal capabilities by incorporating reference audio and event-roll-based editing instructions.

In summary, our main contributions are as follows:

\begin{itemize}

\item We introduce the Audio ControlNet framework for fine-grained text-to-audio (T2A) generation. Without retraining the backbone model, our approach enables controllable T2A generation with respect to loudness, pitch, and sound events.

\item We design two ControlNet variants tailored for fine-grained T2A tasks, namely T2A-ControlNet and T2A-Adapter. Experimental results demonstrate that T2A-Adapter achieves state-of-the-art performance while maintaining a lightweight architecture.

\item We propose representing all control signals as unified temporal sequences to improve extensibility, and design structured representations together with dedicated feature extractors for loudness, pitch, and sound events.

\item We further introduce T2A-Editor, which adopts a ControlNet-like paradigm to transform an existing T2A model into audio editing models that support temporally localized insertion and removal.

\end{itemize}

Our experiments demonstrate that both T2A-ControlNet and T2A-Adapter achieve consistent improvements in controlling loudness, pitch, and sound events. Notably, T2A-Adapter attains state-of-the-art performance on AudioSet-Strong, achieving 54.36 \(F1_{events}\) and 68.26 \(F1_{seg}\) while introducing only 38M additional parameters. Subjective evaluations further indicate that our methods outperform all baseline models \cite{wang2025audiocomposer, hai2024ezaudio}. To facilitate future research, we will release our code, models, and benchmarks.

\section{Related Work}

\subsection{Fine-grained Text-to-audio Generation}

Text-to-audio generation (T2A) aims to generate audio from textual descriptions, with recent progress driven by models like AudioLDM \cite{liu2023audioldm}, Tango \cite{ghosal2023tango}, and AudioBox \cite{vyas2023audiobox}. More recent approaches, including Stable Audio Open \cite{evans2025sao}, EzAudio \cite{hai2024ezaudio}, TangoFlux \cite{hung2024tangoflux}, and MeanAudio \cite{li2025meanaudio}, leverage diffusion transformers \cite{peebles2023dit} to produce high-fidelity, variable-length audio efficiently.

Recently, many studies focus on introducing fine-grained control to T2A models. Early efforts, such as MC-Diffusion \cite{guo2024audiocondition} and PicoAudio \cite{xie2024picoaudio, zheng2025picoaudio2}, train models to handle sound-event-level control, but are limited to fixed event categories or non-overlapping sounds. AudioComposer \cite{wang2025audiocomposer} extends this to open-vocabulary categories by injecting control information in a language-driven form. FreeAudio \cite{jiang2025freeaudio} proposes a training-free approach for fine-grained controllable T2A generation. ControlAudio \cite{jiang2025controlaudio} further expands T2A controllability to speech and employs progressive modeling.

\subsection{Controllable Generation via ControlNet}

Achieving controllability in generative models has been widely studied. In text-to-image generation, ControlNet \cite{zhang2023adding} is a representative approach that introduces control signals through an auxiliary network. Subsequent works such as IP-Adapter \cite{ye2023ip}, T2I-Adapter \cite{mou2024t2i}, Uni-ControlNet \cite{zhao2023unicontrolnet}, and ControlNeXt \cite{peng2024controlnext} extend the ControlNet framework toward more lightweight, efficient, and multi-conditional control.

Similar ideas have also been explored in the audio domain. In the field of music generation, several works \cite{wu2024musiccontrolnet, tsai2025musecontrollite, baker2025lilac, hou2025editing} adapt ControlNet to enable control over musical attributes such as melody, dynamics, chords, and rhythm. More recently, SpecMaskFoley \cite{zhong2025specmaskfoley} applies ControlNet to video-synchronized foley synthesis. 

Despite these advances, fine-grained controllability in text-to-audio generation remains largely underexplored, particularly for control at the sound event level.

\subsection{Fine-grained Audio Editing}

Audio editing aims to modify input audio according to editing instructions, typically expressed as an action-category pair (e.g., ``Add Laughter''). One of the earliest works, AUDIT \cite{wang2023audit}, trains a diffusion model on synthetic data to perform the editing task. However, AUDIT exhibits limited temporal control over when edits occur. Recomposer \cite{ellis2025recomposer} addresses this limitation by incorporating event rolls, enabling edits to be applied precisely at the specified start and end timestamps. While Recomposer aligns closely with our definition of fine-grained audio editing, it still requires training the model from scratch.

\section{FluxAudio}

Before introducing our proposed Audio ControlNet models, we first introduce FluxAudio, the text-to-audio model that serves as the pretrained foundation for this work, originally from \cite{li2025meanaudio}. FluxAudio is built on the FLUX \cite{labs2025flux1kontextflowmatching} architecture and uses MMDiT \cite{esser2024scaling} and DiT \cite{peebles2023dit} as its backbone.

\begin{figure}[htbp]
  \centering
  \includegraphics[width=0.8\columnwidth]{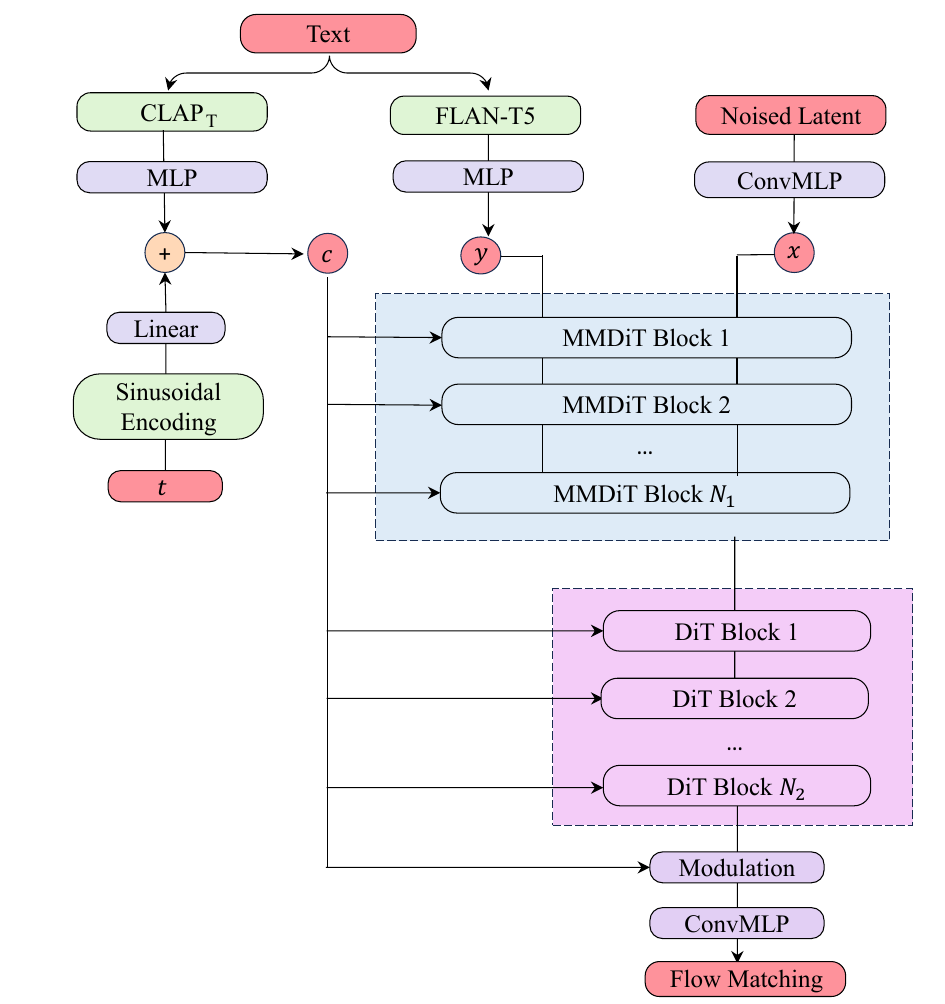}
  \caption{The architecture of FluxAudio, which consists of a hybrid backbone that integrates MMDiT and DiT, and is trained with a flow matching objective.}
  \label{fig:fluxaudio}
\end{figure}

The structure of FluxAudio is illustrated in Figure \ref{fig:fluxaudio}. FluxAudio takes textual inputs encoded by CLAP text encoder \cite{wu2023large} and FLAN-T5 \cite{chung2024scaling}, which are first processed through MLP layers and fused with sinusoidal time embeddings to capture temporal information. The input audio, sampled at 44.1 kHz, is first encoded by a VAE into a latent sequence $x_0$ with frame rate 43.0 Hz and dimension 40. The latent $x_0$ is then noised and passed through $N_1$ MMDiT blocks and $N_2$ DiT blocks, followed by a final ConvMLP to produce the output. FluxAudio is trained with a flow matching objective, defined as:

\begin{equation}
\mathcal{L}_\text{FM} = \mathbb{E}_{x_0, t , \epsilon} \Big\| f_\theta(x_t, t) - \frac{dx_t}{dt} \Big\|^2 
\end{equation}

where $x_t = (1-t)x_0 + t \epsilon$ is the noised audio at time $t$, and $f_\theta$ predicts the velocity field.

\section{Audio ControlNet}

We propose Audio ControlNet, a framework for fine-grained control in text-to-audio generation. The core idea is to introduce an auxiliary conditional network that injects conditions into the pretrained FluxAudio, enabling controllable audio generation without retraining the backbone.

\subsection{ControlNets for Text-to-Audio Generation}
\label{sec:controlnet}

We introduce two variants of Audio ControlNet: T2A-ControlNet and T2A-Adapter, which are inspired by the success of ControlNets \cite{zhang2023adding,ye2023ip} in the text-to-image domain. We adapt them to accommodate the hybrid MMDiT and DiT backbone of FluxAudio, resulting in T2A-ControlNet and T2A-Adapter tailored for text-to-audio models.

\begin{figure*}[htbp]
  \centering
  \includegraphics[width=0.95\linewidth]{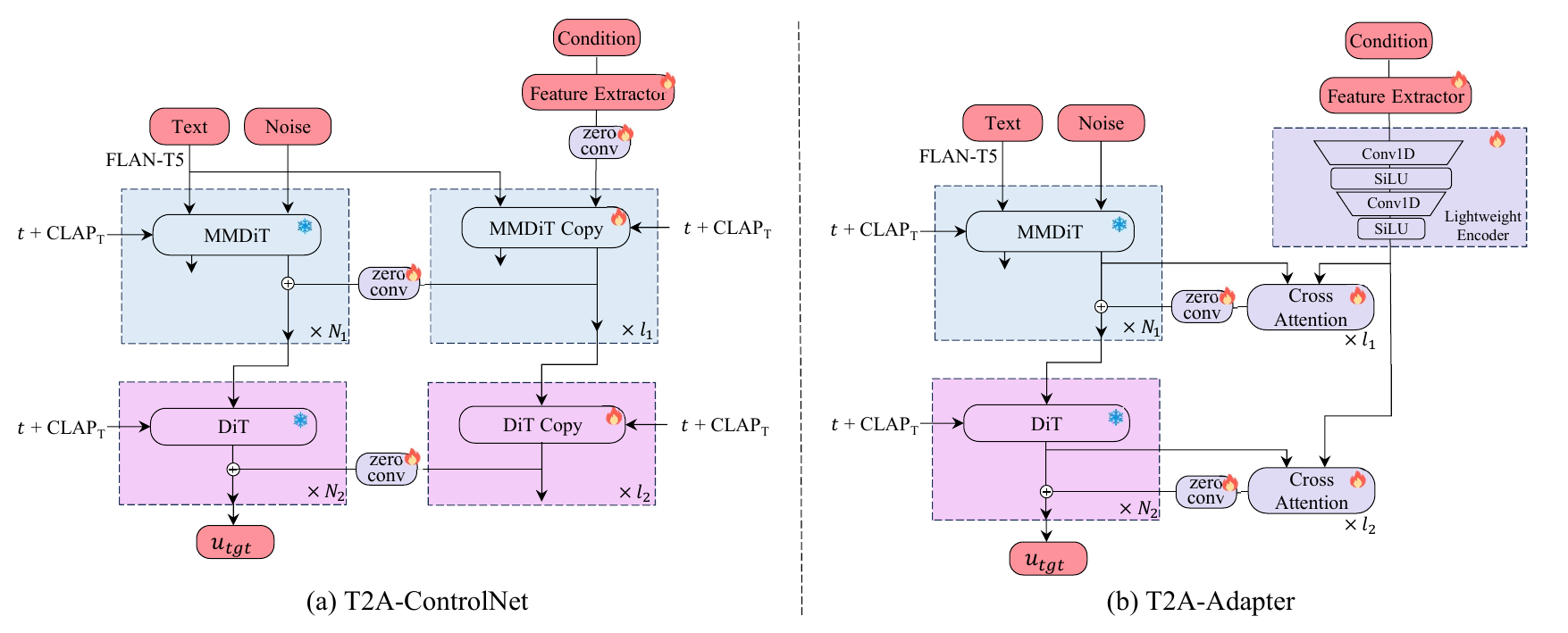}
  \caption{The framework of T2A-ControlNet and T2A-Adapter.}
  \label{fig:framework}
\end{figure*}

\begin{figure}[htbp]
  \centering
  \includegraphics[width=0.95\linewidth]{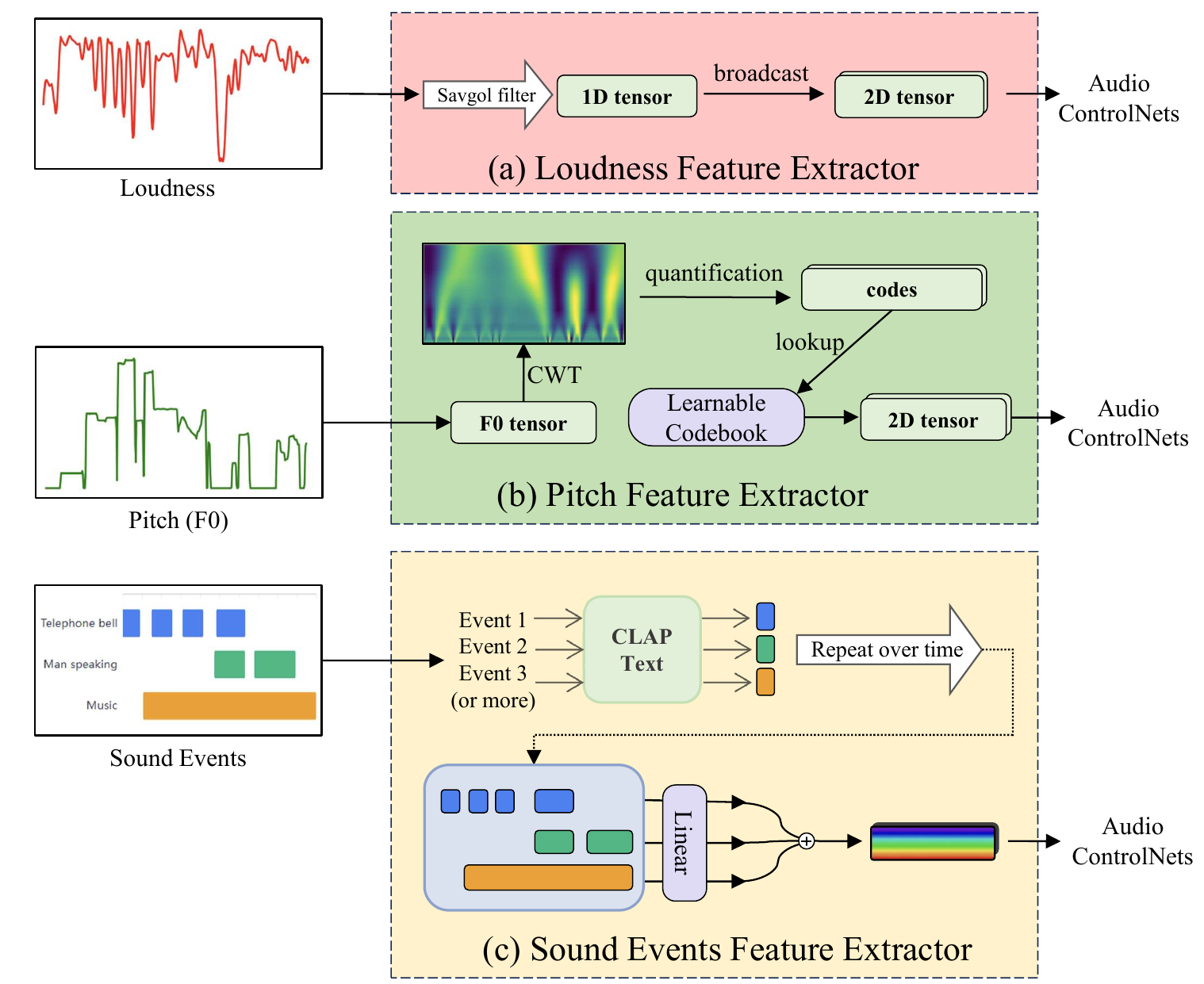}
  \caption{Conditions and their feature extractors.}
  \label{fig:conditions}
\end{figure}

\subsubsection{T2A-ControlNet}

T2A-ControlNet takes time-varying conditions (e.g., loudness, pitch, and sound event roll in Section \ref{sec:controls}) as inputs and provides residual components to the latent representations at each layer of FluxAudio, as illustrated in Figure \ref{fig:framework}(a). Its architecture replicates $l$ layers of FluxAudio, including both MMDiT and DiT layers, where $l$ is a hyperparameter.

Similar to FluxAudio, T2A-ControlNet receives the text prompt and timestep embeddings as inputs; however, its latent input is replaced with the control conditions, with zero-convolution \cite{zhang2023adding} applied at the input. Unlike the standard ControlNet \cite{zhang2023adding}, since FluxAudio uses a hybrid MMDiT and DiT backbone, T2A-ControlNet directly adds the processed latent to each layer in FluxAudio. For DiT layers, the addition is applied directly to the matching latent. For MMDiT layers, the residual is added only to the audio latent components.

During training, the pretrained FluxAudio is frozen, and only the parameters of T2A-ControlNet are updated. This ensures that T2A-ControlNet provides the necessary conditional control information without altering the original FluxAudio.

\subsubsection{T2A-Adapter}

As shown in Figure \ref{fig:framework}(b), T2A-Adapter employs a lightweight encoder to extract features from time-varying conditions. Unlike T2A-ControlNet, T2A-Adapter only takes the control conditions as input.

The lightweight encoder is composed of 1D convolutions and SiLU activation functions \cite{elfwing2018silu}, and it extracts condition features and injects them into the latent representations of the first $l$ layers of FluxAudio through zero-convolution and cross-attention. For DiT layers, the cross-attention operates on the entire latent. For MMDiT layers, the cross-attention is applied only to the audio latent. 

This process can be formally expressed as:
\begin{align}
\mathbf{Q}^{(i)} &= \mathbf{z}^{(i)} \\
[\mathbf{K}^{(i)}, \mathbf{V}^{(i)}] &= \text{LightweightEncoder}(\tilde{\mathbf{c}}^{\mathrm{cond}}) \\
\mathbf{H}^{(i)} &= \text{CrossAttn}(\mathbf{Q}^{(i)}, \mathbf{K}^{(i)}, \mathbf{V}^{(i)}) \\
\mathbf{z}^{(i)}_{\mathrm{out}} &= \mathbf{z}^{(i)} + \text{ZeroConv}(\mathbf{H}^{(i)})
\end{align}
where $\mathbf{z}^{(i)}$ denotes the latent at the $i$-th layer of FluxAudio, serving as the query in cross-attention. The condition $\tilde{\mathbf{c}}^{\mathrm{cond}}$ is processed by the lightweight encoder to produce the key $\mathbf{K}^{(i)}$ and value $\mathbf{V}^{(i)}$.

\subsection{Fine-Grained Conditional Controls}
\label{sec:controls}

In this section, we demonstrate how to inject conditions into Audio ControlNets. We focus on three time-varying conditions, namely loudness, pitch, and sound events. We describe the extraction procedure for each conditioning signal and the feature extractors, as shown in Figure \ref{fig:conditions}.

\subsubsection{Conditions and feature extractors}
\label{sec:conditions}

\paragraph{Loudness}

Loudness is defined as the temporal energy curve of the audio signal. We compute frame-level root-mean-square energy and convert it to the decibel scale. Formally, given an audio signal $x$, we have
\begin{equation}
\mathbf{c}^{\mathrm{loud}}
= 20 \log_{10} \left( E_{\mathrm{rms}}(x) + \epsilon \right),
\end{equation}
where $E_{\mathrm{rms}}(x)$ denotes the RMS energy, and $\epsilon$ is a small constant for numerical stability.

As depicted in \ref{fig:conditions}(a), to obtain a stable loudness control signal, we then apply Savitzky-Golay smoothing \cite{savitzky1964smoothing}:
\begin{equation}
\tilde{\mathbf{c}}^{\mathrm{loud}}
= \mathrm{SavGol}\!\left(\mathbf{c}^{\mathrm{loud}}\right).
\end{equation}

 Once the loudness curve is extracted, it is repeated along the feature dimension and broadcast to match the input dimensionality of the Audio ControlNet. 
 In this way, our loudness feature extractor is implemented in a simple yet effective manner using Savitzky-Golay filtering and broadcasting.

\paragraph{Pitch}

Pitch is defined as the time-varying fundamental frequency of an audio signal. Given an audio signal $x$, we first estimate the continuous fundamental frequency $\mathbf{f}_{0}(x)$ and convert it to the log-frequency domain.

To capture both local and multi-scale variations, we apply a continuous wavelet transform (CWT):
\begin{equation}
\mathbf{c}^{\mathrm{pitch}}_{\mathrm{cwt}} = \mathrm{CWT}\left(\log \mathbf{f}_{0}(x)\right),
\end{equation}
yielding a continuous pitch representation. As shown in Fig.~\ref{fig:conditions}(b), these features are quantized into discrete bins and mapped to the model input through a learnable codebook embedding:
\begin{equation}
\tilde{\mathbf{c}}^{\mathrm{pitch}} = \mathcal{E}\left(\mathrm{Quantize}\left(\mathbf{c}^{\mathrm{pitch}}_{\mathrm{cwt}}\right);\mathcal{Z}\right).
\end{equation}
where $\mathrm{Quantize}(\cdot)$ denotes the pitch quantization operation, and $\mathcal{E}(\cdot;\, \mathcal{Z})$ represents a learnable embedding lookup parameterized by a codebook $\mathcal{Z}$.

\paragraph{Sound Events}
Sound events are represented as an event roll, where the onset and offset times of each of the $N_e$ sound event classes are annotated along the temporal axis, allowing multiple overlapping events (Fig.~\ref{fig:conditions}(c)).

For each event $s_i$, we first extract a semantic embedding from its textual label, using the CLAP text encoder \cite{wu2023large}:
\begin{equation} \mathbf{e}_i = \mathrm{CLAP}_{\mathrm{T}}(s_i), \quad i = 1, \ldots, N_e, \end{equation}

Following the event roll, each embedding is expanded along time, while a zero vector is used when the event is inactive:
\begin{equation}
\mathbf{E}_i(t) =
\begin{cases}
\mathbf{e}_i, & \text{if event } i \text{ is active at time } t, \\
\mathbf{0}, & \text{otherwise}.
\end{cases}
\end{equation}

To aggregate multiple simultaneous events, a shared linear projection $\mathbf{W}$ is applied, and the embeddings are summed across events:
\begin{align} \tilde{\mathbf{c}}^{\mathrm{event}}(t) &= \sum_{i=1}^{N_e} \mathbf{W}\mathbf{E}_i(t) \\ \tilde{\mathbf{c}}^{\mathrm{event}} &= \left[ \tilde{\mathbf{c}}^{\mathrm{event}}(1), \ldots, \tilde{\mathbf{c}}^{\mathrm{event}}(T) \right], \end{align} where $\mathbf{W}$ denotes a learnable linear transformation. The resulting fixed-dimensional temporal representation $\tilde{\mathbf{c}}^{\mathrm{event}}(t)$ is used as the conditioning input to Audio ControlNet models.

\subsection{Audio Editing}

We further extend the Audio ControlNet framework to support audio editing tasks. 
Our motivation is to enable text-to-audio models to perform targeted audio modifications. This extension involves two key aspects: 

\begin{itemize}
    \item First, it should introduce reference audio inputs to the T2A model. This means transforming a text-to-audio model into an audio-to-audio generation model. 
    \item Second, it should incorporate editing instruction as inputs, e.g. ``insert clapping sound: from 2.0 s to 2.5 s'', ``remove people speaking: from 5.0 s to 8.0 s''.
\end{itemize}

\begin{figure}[htbp]
  \centering
  \includegraphics[width=0.9\linewidth]{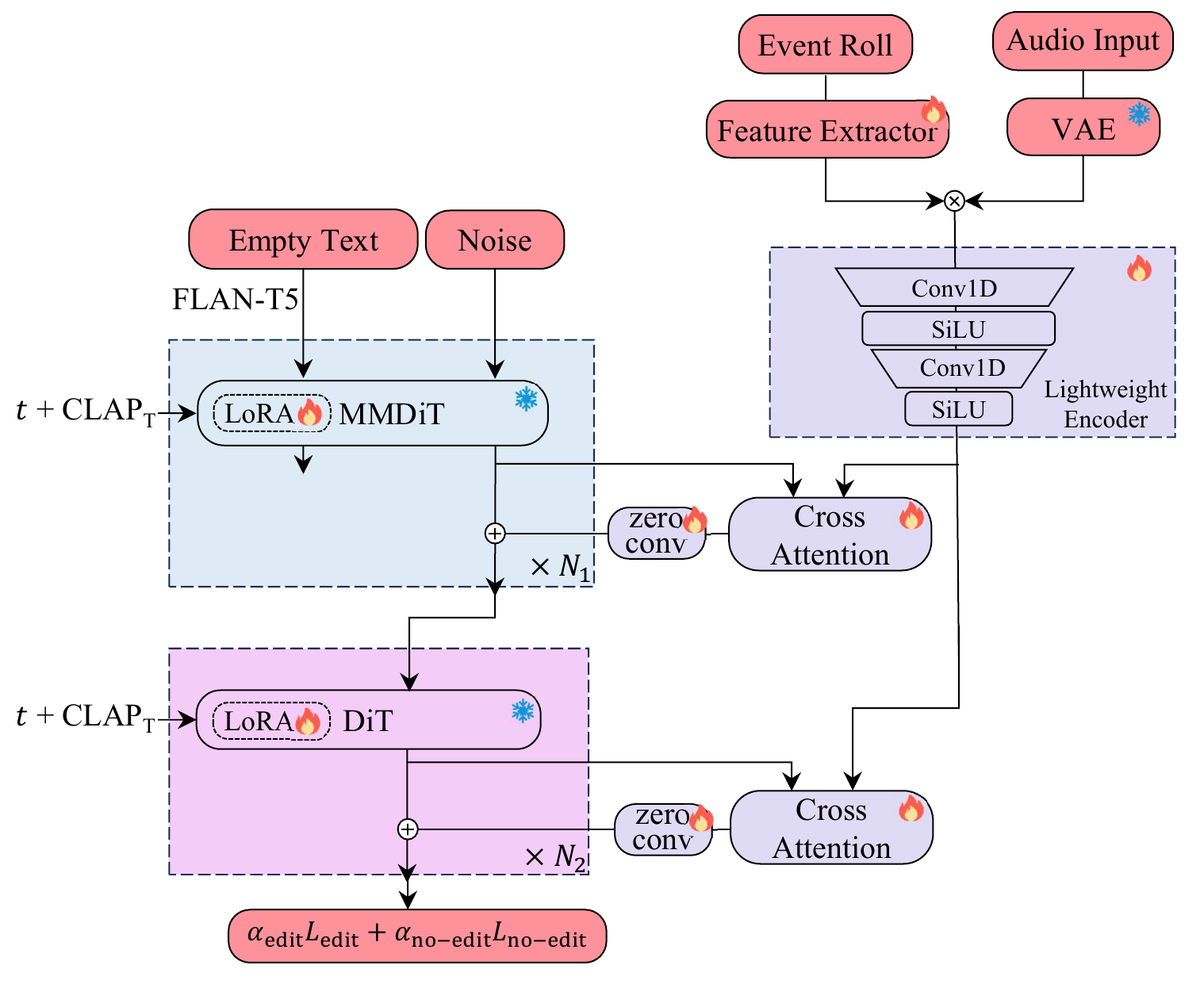}
  \caption{Architecture of T2A-Editor. Reference audio and editing event information are injected via an external network to extend the text-to-audio model for the audio editing task.}
  \label{fig:t2aeditor}
\end{figure}

Taking into account the above aspects, we design the T2A-Editor framework for audio editing, as shown in Figure \ref{fig:t2aeditor}. T2A-Editor adopts an architecture adapted from T2A-Adapter. Specifically, we adopt an additional external conditioning network to inject both the reference audio and editing event information, enabling a text-to-audio model to be seamlessly converted into an audio editing model without modifying its core architecture.

First, we represent editing instructions as an event roll, providing a structured temporal description of target audio events. Textual embeddings are extracted via the CLAP encoder and processed with the same sound event feature pipeline (Section \ref{sec:conditions}) for consistency.

Reference audio VAE latents are concatenated with event features and fed into a lightweight encoder, whose outputs are injected into FluxAudio via cross-attention. T2A-Editor can optionally equip DiT and MMDiT modules with LoRA \cite{hu2022lora} to jointly enhance controllability and audio-to-audio generation quality.

We replace the original text input in FluxAudio with an empty string. During training, a weighted loss emphasizing edited regions ($\alpha_{\text{edit}}:\alpha_{\text{no-edit}} = 10:1$) is employed to encourage learning the desired editing.

\section{Experiments Setup}

\subsection{Datasets}
\label{sec:datasets}

We train our fine-grained control models on the AudioSet-Strong dataset \cite{hershey2021baudiosetstrong}, which contains approximately 80,000 high-quality, manually annotated sound events with precise onset and offset timestamps. We construct training and validation sets with the same data splits as in \cite{guo2024audiocondition}, and report results on the test set.

For audio editing, we use AudioSet-Strong as background audio and FSD50K \cite{fonseca2021fsd50k} as target audio. We first apply an energy threshold to filter FSD50K and segment the audio, retaining clips between 0.5 and 4 seconds. Target clips are then randomly inserted into background audio at random positions, while avoiding overlap between the target class and the audio caption. 

\subsection{Model Configuration}

FluxAudio comprises $N_1 = 4$ MMDiT layers and $N_2 = 8$ DiT layers (460M parameters), generating 10-second audio clips at 44.1 kHz, encoded by the VAE into a latent dimension of 40 at 43.0 Hz. It is pre-trained on 7.8k hours of audio from AudioCaps \cite{kim2019audiocaps}, AudioSet \cite{gemmeke2017audioset, bai2025audiosetcaps}, FreeSound \cite{mei2024wavcaps}, and other datasets \cite{chen2020vggsound, doh2023lp, hershey2021baudiosetstrong}. We ensure that there is no overlap with the test set in Section~\ref{sec:datasets}.

For the ControlNet depth hyperparameter $l$, we experiment with $l \in \{4, 8, 12\}$ and report the best results among these configurations (see Section~\ref{sec:effect_depth}).  

Loudness features are computed as RMS energy with frame length 4096, hop 1025 (43.0 fps), smoothed via Savitzky-Golay (window 11, poly order 3). Pitch features are extracted using pYIN and Ricker wavelets over scales 1-32, quantized into 256 bins.

The lightweight encoder has 3 layers of 1D convolution with SiLU, with only 0.77M parameters. LoRA is injected into T2A-Editor with rank 64.

\subsection{Training and Inference}

For the Audio ControlNets conditioned on loudness, pitch, and sound events, we use a batch size of 64 and a learning rate of $1\times10^{-4}$. For the insert and remove editing models, all paired training data are extracted online during training, and a larger batch size of 128 is adopted. All models are optimized using AdamW and trained for 120k steps on four GPUs, each with 48 GB memory.

We adopt classifier-free guidance (CFG) by treating the caption input and the control conditions as a unified conditioning signal. During training, caption and control conditions are simultaneously dropped with a probability of 10\%. We apply CFG scale 4.5 and 25 flow-matching steps at inference.

\subsection{Evaluation Metrics}

\paragraph{Objective metrics.} For loudness and pitch control, we evaluate the generated audio by computing the mean absolute error (MAE) between the generated and reference features. For sound event control, we employ PB-SED \cite{ebbers2022pbsed} as the sound event detection model and report both the event-based F1-score ($F_{1}^{\text{event}}$) and the segment-based F1-score ($F_{1}^{\text{seg}}$) \cite{mesaros2016metrics} to assess event accuracy.

For fine-grained audio editing, we found that sequence-level metrics often fail to accurately capture fine-grained edits, as they require precise temporal localization. To address this, we use FlexSED \cite{hai2025flexsed}, a language-query sound event detection model. The target class name (for insertion or removal) is fed into FlexSED, and the mean output over the edited time interval is computed to measure the correctness of the operation. 
Higher scores indicate successful insertion, while lower scores indicate successful removal.

\paragraph{Subjective metrics.}

We evaluate the controllability of our model using a mean opinion score (MOS) protocol from 3 aspects. \(MOS_\text{loud}\) measures how accurately the generated audio follows the target loudness. \(MOS_\text{pitch}\) evaluates the correctness of pitch control. \(MOS_\text{events}\) assesses the consistency between generated sound events and the specified event roll.
For each task, 5 test groups from each model are rated by 20 evaluators, and the mean score is calculated.

\section{Results and Analysis}

We compare our proposed T2A-ControlNet and T2A-Adapter with state-of-the-art text-to-audio models, including TangoFlux \cite{hung2024tangoflux}, Stable Audio Open \cite{evans2025sao}, EzAudio \cite{hai2024ezaudio}, and FluxAudio \cite{li2025meanaudio}. For loudness control, we additionally compare with the energy ControlNet provided by the EzAudio repository\footnote{https://github.com/haidog-yaqub/EzAudio}. For sound event control, we include AudioComposer \cite{wang2025audiocomposer} as a baseline. For audio editing, since existing related methods \cite{ellis2025recomposer} do not provide publicly available implementations, we compare the edited results with the input audio and the ground-truth target.

\subsection{Loudness and Pitch Control}

As shown in Table~\ref{tab:loudness_pitch_res}, our proposed methods achieve superior control over both loudness and pitch. Specifically, T2A-Adapter attains an MAE of 1.40 on loudness using only 38M trainable parameters. T2A-ControlNet achieves better pitch performance (MAE 119.28) but requires a much larger parameter count (410M). Compared to FluxAudio, the T2A-Adapter offers a clear improvement while maintaining a compact model size. Moreover, T2A-Adapter outperforms EzAudio-L-Control on the loudness metric, demonstrating strong loudness control capabilities. Subjective metrics also demonstrate the effectiveness of T2A-ControlNet and T2A-Adapter.

\setlength{\tabcolsep}{1mm}
\begin{table}[htbp]
    \centering
    \caption{Text-to-audio results under loudness and pitch control.}
    \resizebox{1\linewidth}{!}{
    \begin{tabular}{l|c|c|c|c|c}
        \toprule[1.5pt]
             \multirow{2}{*}{\textbf{Model}} & \multirow{2}{*}{\begin{tabular}[c]{@{}c@{}}\textbf{Trainable} \\ \textbf{Params}\end{tabular}}  &   \multicolumn{1}{c|}{\textbf{Loudness}} & \multicolumn{1}{c}{\textbf{Pitch  }} & \multicolumn{2}{|c}{\textbf{Subjective }}  \\
             & & MAE $\downarrow$ & MAE $\downarrow$ & \(MOS_\text{loud}\)$\uparrow$ & \(MOS_\text{pitch}\)$\uparrow$\\
        \hline 
            Ground Truth & - & - & - & 3.47 & 3.18 \\
            \hline
            TangoFlux & 819M & 11.41 & 239.46 & - & - \\
            Stable Audio Open  &  1050M  & 17.49 &  233.88 & - & - \\
            EzAudio-XL & 875M &  11.03  & 203.10 & 1.99 & 1.77 \\
            EzAudio-L-Energy & 278M & 2.22 & - & 3.28 & - \\
            FluxAudio & 460M & 13.36 & 251.89 & 2.81 & 1.94  \\
            \hline
            T2A-ControlNet & 410M & 4.08 & \textbf{119.28} & \textbf{3.65} & \textbf{3.04} \\
            T2A-Adapter & 38M & \textbf{1.40} &  148.02 & 3.64 & \textbf{3.04} \\
        \bottomrule[1.5pt]
    \end{tabular}
    }
    \label{tab:loudness_pitch_res}
\end{table}

\subsection{Sound Events Control}

As shown in Table~\ref{tab:events_res}, our proposed method demonstrates strong performance on sound event control. In particular, T2A-Adapter achieves the best results with an \(F1_{event}\) score of 54.36 and an \(F1_{seg}\) score of 68.26, outperforming all baseline text-to-audio models. Compared to FluxAudio, T2A-Adapter yields a substantial improvement in both event-level and segment-level metrics, while introducing only 38M trainable parameters. 

Furthermore, T2A-Adapter consistently surpasses T2A-ControlNet, achieving higher control accuracy with significantly fewer trainable parameters. Notably, T2A-Adapter also outperforms AudioComposer-S and AudioComposer-L, indicating superior sound event controllability when compared with dedicated models. 

Subjective evaluations show that both T2A-Adapter and T2A-ControlNet outperform all baseline methods, with T2A-Adapter performing slightly better. On \(MOS_\text{events}\), our models even achieve scores above the ground truth, possibly reflecting the benefit of the backbone pretrained on large-scale datasets, which may also enlarge the performance of Audio ControlNet.

\setlength{\tabcolsep}{1mm}
\begin{table}[htbp]
    \centering
    \caption{Text-to-audio results under sound events control. Results for AudioComposer-L are taken directly from the paper \cite{wang2025audiocomposer}, as its weights are not publicly available.}
    \resizebox{1\linewidth}{!}{
    \begin{tabular}{l|c|cc|cc|c}
        \toprule[1.5pt]
             \multirow{2}{*}{\textbf{Model}} & \multirow{2}{*}{\begin{tabular}[c]{@{}c@{}}\textbf{Trainable} \\ \textbf{Params}\end{tabular}}  &   \multicolumn{2}{c|}{\textbf{PB-SED}} & \textbf{Subjective}\ \\
             & & \(F1_{event}\) $\uparrow$ & \(F1_{seg}\) $\uparrow$ &  \(MOS_\text{events}\) $\uparrow$ \\
            \hline 
            Ground Truth & - & 43.36 & 63.46 & 3.21 \\
            \hline 
            TangoFlux & 820M & 6.29 & 43.14 & - \\
            Stable Audio Open  &  1050M & 6.05 & 31.62 & - \\
            EzAudio & 875M & 10.43 & 50.18 & 2.16 \\
            FluxAudio & 460M & 10.01 & 50.59 & 2.05 \\
            \hline 
            AudioComposer-S & 272M & 43.51 & 60.83 & 2.84 \\
            AudioComposer-L & 743M & 44.40* & 63.30* & - \\
            \hline
            T2A-ControlNet & 410M & 47.98 & 67.92 & 3.71 \\
            T2A-Adapter & 38M & \textbf{54.36} & \textbf{68.26} & \textbf{3.76} \\
        \bottomrule[1.5pt]
    \end{tabular}
    }
    \label{tab:events_res}
\end{table}


\subsection{Multi-condition Control}

Although Audio ControlNets with different conditions are trained independently, they can be composed at inference time by directly summing their latent representations to enable multi-condition control\footnote{Additional examples of multi-condition control are available at the demo link.}.
This is enabled by the strong compositionality of ControlNet as discussed in \cite{zhang2023adding}, which allows multiple conditions to be jointly applied in a straight forward manner.

\subsection{Audio Editing}

As shown in Table~\ref{tab:editing_res}, we evaluate fine-grained text-to-audio editing performance under both insertion and removal settings using the FlexSED metric. Compared to the input audio, T2A-Editor substantially improves editing quality in both scenarios, indicating effective text-guided audio manipulation.

In particular, incorporating LoRA further enhances performance, with T2A-Editor w/ LoRA achieving a FlexSED score of 0.1340 for insertion and 0.0429 for removal, outperforming the vanilla T2A-Editor in both cases. Notably, the edited results move significantly closer to the ground-truth upper bound, demonstrating that our method can work with only a small number of additional trainable parameters.

\setlength{\tabcolsep}{1mm}
\begin{table}[htbp]
    \centering
    \caption{Fine-grained text-to-audio editing results.}
    \resizebox{\linewidth}{!}{
    \begin{tabular}{l|c|cc|cc}
        \toprule[1.5pt]
             \multirow{2}{*}{\textbf{Model}} & \multirow{2}{*}{\begin{tabular}[c]{@{}c@{}}\textbf{Trainable} \\ \textbf{Params}\end{tabular}}  &   \multicolumn{2}{c|}{\textbf{Insert}} & \multicolumn{2}{c}{\textbf{Remove}}  \\
             & & FlexSED$\uparrow$ & & FlexSED$\downarrow$ &  \\
        \hline 
            Input & - & 0.0257 & & 0.1946 & \\
            Ground Truth & - & 0.1946 & & 0.0257 & \\
            \hline
            T2A-Editor & 14M & 0.0989 & & 0.0679\\
            T2A-Editor w/ LoRA & 14M+4M & \textbf{0.1340} & & \textbf{0.0429}\\
        \bottomrule[1.5pt]
    \end{tabular}
    }
    \label{tab:editing_res}
\end{table}

\subsection{Effect of ControlNet Depth}
\label{sec:effect_depth}

\begin{figure}[h]
    \centering  \includegraphics[width=0.8\linewidth]{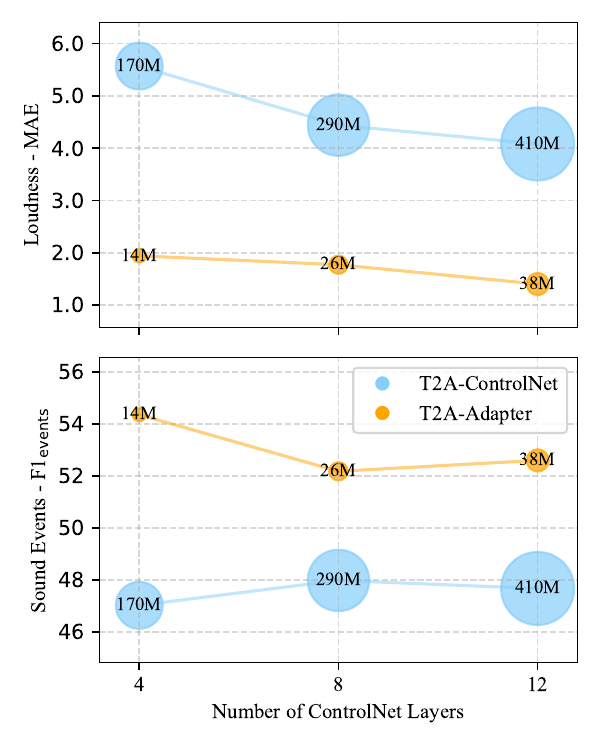}
  \caption{Effect of ControlNet layers on performance. Marker size denotes models' trainable parameters.}
  \label{fig:control_depth}
\end{figure}

Figure~\ref{fig:control_depth} analyzes the effect of ControlNet depth on both loudness control and sound events control for T2A-ControlNet and T2A-Adapter. As the depth increases, T2A-ControlNet exhibits an improvement in loudness MAE, decreasing from 5.6 at depth 4 to 4.0 at depth 12. However, the performance on sound events control remains largely saturated, with only marginal variation in \(F1_{event}\).

For the T2A-Adapter, increasing the depth leads to a clear improvement in loudness control. For sound events control, T2A-Adapter achieves high \(F1_{event}\) scores even at shallow depth, indicating its effectiveness.

Overall, these results suggest that T2A-Adapter is less sensitive to ControlNet depth and can achieve strong control performance with a shallower depth, highlighting its superior parameter efficiencycompared to T2A-ControlNet.


\section{Conclusion}

In this work, we introduce Audio ControlNet, a framework for fine-grained text-to-audio generation and editing that augments pre-trained T2A models with lightweight control networks. Our designs, T2A-ControlNet and T2A-Adapter, enable precise control over loudness, pitch, and sound events. T2A-Adapter can achieve strong performance while using fewer parameters than T2A-ControlNet. We further extended this approach to audio editing with T2A-Editor, supporting temporally localized insertion and removal of audio events. Overall, our results demonstrate that Audio ControlNet models can provide precise, extensible control and editing for T2A models.

\section*{Limitations}
This work has several limitations. Due to computational constraints, we do not exhaustively explore hyperparameters, and a finer search may yield lighter or stronger Audio ControlNets. 
Multi-condition control in our framework mainly relies on generalization at inference time. Incorporating joint multi-condition supervision during training may further improve multi-condition controllability. Finally, we only study a limited set of conditions; richer control signals, such as speech emotion or singing melody, are left for future work.



\bibliography{custom}

@article{li2025meanaudio,
  title={MeanAudio: Fast and Faithful Text-to-Audio Generation with Mean Flows},
  author={Li, Xiquan and Liu, Junxi and Liang, Yuzhe and Niu, Zhikang and Chen, Wenxi and Chen, Xie},
  journal={arXiv preprint arXiv:2508.06098},
  year={2025}
}

@inproceedings{esser2024scaling,
  title={Scaling rectified flow transformers for high-resolution image synthesis},
  author={Esser, Patrick and Kulal, Sumith and Blattmann, Andreas and Entezari, Rahim and M{\"u}ller, Jonas and Saini, Harry and Levi, Yam and Lorenz, Dominik and Sauer, Axel and Boesel, Frederic and others},
  booktitle={Forty-first international conference on machine learning},
  year={2024}
}

@misc{labs2025flux1kontextflowmatching,
      title={FLUX.1 Kontext: Flow Matching for In-Context Image Generation and Editing in Latent Space},
      author={Black Forest Labs and Stephen Batifol and Andreas Blattmann and Frederic Boesel and Saksham Consul and Cyril Diagne and Tim Dockhorn and Jack English and Zion English and Patrick Esser and Sumith Kulal and Kyle Lacey and Yam Levi and Cheng Li and Dominik Lorenz and Jonas Müller and Dustin Podell and Robin Rombach and Harry Saini and Axel Sauer and Luke Smith},
      year={2025},
      eprint={2506.15742},
      archivePrefix={arXiv},
      primaryClass={cs.GR},
      url={https://arxiv.org/abs/2506.15742},
}

@inproceedings{zhang2023adding,
  title={Adding conditional control to text-to-image diffusion models},
  author={Zhang, Lvmin and Rao, Anyi and Agrawala, Maneesh},
  booktitle={Proceedings of the IEEE/CVF international conference on computer vision},
  pages={3836--3847},
  year={2023}
}

@article{ye2023ip,
  title={Ip-adapter: Text compatible image prompt adapter for text-to-image diffusion models},
  author={Ye, Hu and Zhang, Jun and Liu, Sibo and Han, Xiao and Yang, Wei},
  journal={arXiv preprint arXiv:2308.06721},
  year={2023}
}

@article{elfwing2018silu,
  title={Sigmoid-weighted linear units for neural network function approximation in reinforcement learning},
  author={Elfwing, Stefan and Uchibe, Eiji and Doya, Kenji},
  journal={Neural networks},
  volume={107},
  pages={3--11},
  year={2018},
  publisher={Elsevier}
}

@article{savitzky1964smoothing,
  title={Smoothing and differentiation of data by simplified least squares procedures.},
  author={Savitzky, Abraham and Golay, Marcel JE},
  journal={Analytical chemistry},
  volume={36},
  number={8},
  pages={1627--1639},
  year={1964},
  publisher={ACS Publications}
}

@inproceedings{wu2023large,
  title={Large-scale contrastive language-audio pretraining with feature fusion and keyword-to-caption augmentation},
  author={Wu, Yusong and Chen, Ke and Zhang, Tianyu and Hui, Yuchen and Berg-Kirkpatrick, Taylor and Dubnov, Shlomo},
  booktitle={ICASSP 2023-2023 IEEE International Conference on Acoustics, Speech and Signal Processing (ICASSP)},
  pages={1--5},
  year={2023},
  organization={IEEE}
}

@article{hu2022lora,
  title={Lora: Low-rank adaptation of large language models.},
  author={Hu, Edward J and Shen, Yelong and Wallis, Phillip and Allen-Zhu, Zeyuan and Li, Yuanzhi and Wang, Shean and Wang, Lu and Chen, Weizhu and others},
  journal={ICLR},
  volume={1},
  number={2},
  pages={3},
  year={2022}
}

@inproceedings{evans2025sao,
  title={Stable audio open},
  author={Evans, Zach and Parker, Julian D and Carr, CJ and Zukowski, Zack and Taylor, Josiah and Pons, Jordi},
  booktitle={ICASSP 2025-2025 IEEE International Conference on Acoustics, Speech and Signal Processing (ICASSP)},
  pages={1--5},
  year={2025},
  organization={IEEE}
}

@article{liu2023audioldm,
  title={Audioldm: Text-to-audio generation with latent diffusion models},
  author={Liu, Haohe and Chen, Zehua and Yuan, Yi and Mei, Xinhao and Liu, Xubo and Mandic, Danilo and Wang, Wenwu and Plumbley, Mark D},
  journal={arXiv preprint arXiv:2301.12503},
  year={2023}
}

@article{copet2023musicgen,
  title={Simple and controllable music generation},
  author={Copet, Jade and Kreuk, Felix and Gat, Itai and Remez, Tal and Kant, David and Synnaeve, Gabriel and Adi, Yossi and D{\'e}fossez, Alexandre},
  journal={Advances in Neural Information Processing Systems},
  volume={36},
  pages={47704--47720},
  year={2023}
}

@inproceedings{cheng2025mmaudio,
  title={MMAudio: Taming Multimodal Joint Training for High-Quality Video-to-Audio Synthesis},
  author={Cheng, Ho Kei and Ishii, Masato and Hayakawa, Akio and Shibuya, Takashi and Schwing, Alexander and Mitsufuji, Yuki},
  booktitle={Proceedings of the Computer Vision and Pattern Recognition Conference},
  pages={28901--28911},
  year={2025}
}

@article{liu2024audioldm2,
  title={Audioldm 2: Learning holistic audio generation with self-supervised pretraining},
  author={Liu, Haohe and Yuan, Yi and Liu, Xubo and Mei, Xinhao and Kong, Qiuqiang and Tian, Qiao and Wang, Yuping and Wang, Wenwu and Wang, Yuxuan and Plumbley, Mark D},
  journal={IEEE/ACM Transactions on Audio, Speech, and Language Processing},
  volume={32},
  pages={2871--2883},
  year={2024},
  publisher={IEEE}
}

@article{hai2024ezaudio,
  title={Ezaudio: Enhancing text-to-audio generation with efficient diffusion transformer},
  author={Hai, Jiarui and Xu, Yong and Zhang, Hao and Li, Chenxing and Wang, Helin and Elhilali, Mounya and Yu, Dong},
  journal={arXiv preprint arXiv:2409.10819},
  year={2024}
}

@inproceedings{wang2025audiocomposer,
  title={AudioComposer: Towards Fine-grained Audio Generation with Natural Language Descriptions},
  author={Wang, Yuanyuan and Chen, Hangting and Yang, Dongchao and Wu, Zhiyong and Wu, Xixin},
  booktitle={ICASSP 2025-2025 IEEE International Conference on Acoustics, Speech and Signal Processing (ICASSP)},
  pages={1--5},
  year={2025},
  organization={IEEE}
}

@article{xie2024picoaudio,
  title={Picoaudio: Enabling precise timestamp and frequency controllability of audio events in text-to-audio generation},
  author={Xie, Zeyu and Xu, Xuenan and Wu, Zhizheng and Wu, Mengyue},
  journal={arXiv preprint arXiv:2407.02869},
  year={2024}
}

@article{zheng2025picoaudio2,
  title={PicoAudio2: Temporal Controllable Text-to-Audio Generation with Natural Language Description},
  author={Zheng, Zihao and Xie, Zeyu and Xu, Xuenan and Wu, Wen and Zhang, Chao and Wu, Mengyue},
  journal={arXiv preprint arXiv:2509.00683},
  year={2025}
}

@article{jiang2025controlaudio,
  title={ControlAudio: Tackling Text-Guided, Timing-Indicated and Intelligible Audio Generation via Progressive Diffusion Modeling},
  author={Jiang, Yuxuan and Chen, Zehua and Ju, Zeqian and Dai, Yusheng and Dou, Weibei and Zhu, Jun},
  journal={arXiv preprint arXiv:2510.08878},
  year={2025}
}

@inproceedings{mou2024t2i,
  title={T2i-adapter: Learning adapters to dig out more controllable ability for text-to-image diffusion models},
  author={Mou, Chong and Wang, Xintao and Xie, Liangbin and Wu, Yanze and Zhang, Jian and Qi, Zhongang and Shan, Ying},
  booktitle={Proceedings of the AAAI conference on artificial intelligence},
  volume={38},
  number={5},
  pages={4296--4304},
  year={2024}
}

@article{zhao2023unicontrolnet,
  title={Uni-controlnet: All-in-one control to text-to-image diffusion models},
  author={Zhao, Shihao and Chen, Dongdong and Chen, Yen-Chun and Bao, Jianmin and Hao, Shaozhe and Yuan, Lu and Wong, Kwan-Yee K},
  journal={Advances in Neural Information Processing Systems},
  volume={36},
  pages={11127--11150},
  year={2023}
}

@inproceedings{hershey2021baudiosetstrong,
  title={The benefit of temporally-strong labels in audio event classification},
  author={Hershey, Shawn and Ellis, Daniel PW and Fonseca, Eduardo and Jansen, Aren and Liu, Caroline and Moore, R Channing and Plakal, Manoj},
  booktitle={ICASSP 2021-2021 IEEE International Conference on Acoustics, Speech and Signal Processing (ICASSP)},
  pages={366--370},
  year={2021},
  organization={IEEE}
}

@inproceedings{guo2024audiocondition,
  title={Audio generation with multiple conditional diffusion model},
  author={Guo, Zhifang and Mao, Jianguo and Tao, Rui and Yan, Long and Ouchi, Kazushige and Liu, Hong and Wang, Xiangdong},
  booktitle={Proceedings of the AAAI Conference on Artificial Intelligence},
  volume={38},
  number={16},
  pages={18153--18161},
  year={2024}
}

@article{ebbers2022pbsed,
  title={Pre-training and self-training for sound event detection in domestic environments},
  author={Ebbers, Janek and Haeb-Umbach, Reinhold},
  year={2022}
}

@article{mesaros2016metrics,
  title={Metrics for polyphonic sound event detection},
  author={Mesaros, Annamaria and Heittola, Toni and Virtanen, Tuomas},
  journal={Applied Sciences},
  volume={6},
  number={6},
  pages={162},
  year={2016},
  publisher={MDPI}
}

@article{hung2024tangoflux,
  title={Tangoflux: Super fast and faithful text to audio generation with flow matching and clap-ranked preference optimization},
  author={Hung, Chia-Yu and Majumder, Navonil and Kong, Zhifeng and Mehrish, Ambuj and Bagherzadeh, Amir Ali and Li, Chuan and Valle, Rafael and Catanzaro, Bryan and Poria, Soujanya},
  journal={arXiv preprint arXiv:2412.21037},
  year={2024}
}

@article{ellis2025recomposer,
  title={Recomposer: Event-roll-guided generative audio editing},
  author={Ellis, Daniel PW and Fonseca, Eduardo and Weiss, Ron J and Wilson, Kevin and Wisdom, Scott and Erdogan, Hakan and Hershey, John R and Jansen, Aren and Moore, R Channing and Plakal, Manoj},
  journal={arXiv preprint arXiv:2509.05256},
  year={2025}
}

@inproceedings{ghosal2023tango,
  title={Text-to-audio generation using instruction guided latent diffusion model},
  author={Ghosal, Deepanway and Majumder, Navonil and Mehrish, Ambuj and Poria, Soujanya},
  booktitle={Proceedings of the 31st ACM International Conference on Multimedia},
  pages={3590--3598},
  year={2023}
}

@article{vyas2023audiobox,
  title={Audiobox: Unified audio generation with natural language prompts},
  author={Vyas, Apoorv and Shi, Bowen and Le, Matthew and Tjandra, Andros and Wu, Yi-Chiao and Guo, Baishan and Zhang, Jiemin and Zhang, Xinyue and Adkins, Robert and Ngan, William and others},
  journal={arXiv preprint arXiv:2312.15821},
  year={2023}
}

@inproceedings{peebles2023dit,
  title={Scalable diffusion models with transformers},
  author={Peebles, William and Xie, Saining},
  booktitle={Proceedings of the IEEE/CVF international conference on computer vision},
  pages={4195--4205},
  year={2023}
}

@article{peng2024controlnext,
  title={Controlnext: Powerful and efficient control for image and video generation},
  author={Peng, Bohao and Wang, Jian and Zhang, Yuechen and Li, Wenbo and Yang, Ming-Chang and Jia, Jiaya},
  journal={arXiv preprint arXiv:2408.06070},
  year={2024}
}

@article{wu2024musiccontrolnet,
  title={Music controlnet: Multiple time-varying controls for music generation},
  author={Wu, Shih-Lun and Donahue, Chris and Watanabe, Shinji and Bryan, Nicholas J},
  journal={IEEE/ACM Transactions on Audio, Speech, and Language Processing},
  volume={32},
  pages={2692--2703},
  year={2024},
  publisher={IEEE}
}

@inproceedings{hou2025editing,
  title={Editing music with melody and text: Using controlnet for diffusion transformer},
  author={Hou, Siyuan and Liu, Shansong and Yuan, Ruibin and Xue, Wei and Shan, Ying and Zhao, Mangsuo and Zhang, Chao},
  booktitle={ICASSP 2025-2025 IEEE International Conference on Acoustics, Speech and Signal Processing (ICASSP)},
  pages={1--5},
  year={2025},
  organization={IEEE}
}

@article{baker2025lilac,
  title={LiLAC: A Lightweight Latent ControlNet for Musical Audio Generation},
  author={Baker, Tom and Nistal, Javier},
  journal={arXiv preprint arXiv:2506.11476},
  year={2025}
}

@article{tsai2025musecontrollite,
  title={MuseControlLite: Multifunctional Music Generation with Lightweight Conditioners},
  author={Tsai, Fang-Duo and Wu, Shih-Lun and Lee, Weijaw and Yang, Sheng-Ping and Chen, Bo-Rui and Cheng, Hao-Chung and Yang, Yi-Hsuan},
  journal={arXiv preprint arXiv:2506.18729},
  year={2025}
}

@article{zhong2025specmaskfoley,
  title={SpecMaskFoley: Steering Pretrained Spectral Masked Generative Transformer Toward Synchronized Video-to-audio Synthesis via ControlNet},
  author={Zhong, Zhi and Takahashi, Akira and Cui, Shuyang and Toyama, Keisuke and Takahashi, Shusuke and Mitsufuji, Yuki},
  journal={arXiv preprint arXiv:2505.16195},
  year={2025}
}

@inproceedings{jiang2025freeaudio,
  title={Freeaudio: Training-free timing planning for controllable long-form text-to-audio generation},
  author={Jiang, Yuxuan and Chen, Zehua and Ju, Zeqian and Li, Chang and Dou, Weibei and Zhu, Jun},
  booktitle={Proceedings of the 33rd ACM International Conference on Multimedia},
  pages={9871--9880},
  year={2025}
}

@article{wang2023audit,
  title={Audit: Audio editing by following instructions with latent diffusion models},
  author={Wang, Yuancheng and Ju, Zeqian and Tan, Xu and He, Lei and Wu, Zhizheng and Bian, Jiang and others},
  journal={Advances in Neural Information Processing Systems},
  volume={36},
  pages={71340--71357},
  year={2023}
}

@article{chung2024scaling,
  title={Scaling instruction-finetuned language models},
  author={Chung, Hyung Won and Hou, Le and Longpre, Shayne and Zoph, Barret and Tay, Yi and Fedus, William and Li, Yunxuan and Wang, Xuezhi and Dehghani, Mostafa and Brahma, Siddhartha and others},
  journal={Journal of Machine Learning Research},
  volume={25},
  number={70},
  pages={1--53},
  year={2024}
}

@inproceedings{kim2019audiocaps,
  title={Audiocaps: Generating captions for audios in the wild},
  author={Kim, Chris Dongjoo and Kim, Byeongchang and Lee, Hyunmin and Kim, Gunhee},
  booktitle={Proceedings of the 2019 Conference of the North American Chapter of the Association for Computational Linguistics: Human Language Technologies, Volume 1 (Long and Short Papers)},
  pages={119--132},
  year={2019}
}

@inproceedings{gemmeke2017audioset,
  title={Audio set: An ontology and human-labeled dataset for audio events},
  author={Gemmeke, Jort F and Ellis, Daniel PW and Freedman, Dylan and Jansen, Aren and Lawrence, Wade and Moore, R Channing and Plakal, Manoj and Ritter, Marvin},
  booktitle={2017 IEEE international conference on acoustics, speech and signal processing (ICASSP)},
  pages={776--780},
  year={2017},
  organization={IEEE}
}

@article{bai2025audiosetcaps,
  title={Audiosetcaps: An enriched audio-caption dataset using automated generation pipeline with large audio and language models},
  author={Bai, Jisheng and Liu, Haohe and Wang, Mou and Shi, Dongyuan and Wang, Wenwu and Plumbley, Mark D and Gan, Woon-Seng and Chen, Jianfeng},
  journal={IEEE Transactions on Audio, Speech and Language Processing},
  year={2025},
  publisher={IEEE}
}

@article{mei2024wavcaps,
  title={Wavcaps: A chatgpt-assisted weakly-labelled audio captioning dataset for audio-language multimodal research},
  author={Mei, Xinhao and Meng, Chutong and Liu, Haohe and Kong, Qiuqiang and Ko, Tom and Zhao, Chengqi and Plumbley, Mark D and Zou, Yuexian and Wang, Wenwu},
  journal={IEEE/ACM Transactions on Audio, Speech, and Language Processing},
  volume={32},
  pages={3339--3354},
  year={2024},
  publisher={IEEE}
}

@inproceedings{chen2020vggsound,
  title={Vggsound: A large-scale audio-visual dataset},
  author={Chen, Honglie and Xie, Weidi and Vedaldi, Andrea and Zisserman, Andrew},
  booktitle={ICASSP 2020-2020 IEEE International Conference on Acoustics, Speech and Signal Processing (ICASSP)},
  pages={721--725},
  year={2020},
  organization={IEEE}
}

@article{doh2023lp,
  title={Lp-musiccaps: Llm-based pseudo music captioning},
  author={Doh, SeungHeon and Choi, Keunwoo and Lee, Jongpil and Nam, Juhan},
  journal={arXiv preprint arXiv:2307.16372},
  year={2023}
}

@article{hai2025flexsed,
  title={FlexSED: Towards Open-Vocabulary Sound Event Detection},
  author={Hai, Jiarui and Wang, Helin and Guo, Weizhe and Elhilali, Mounya},
  journal={arXiv preprint arXiv:2509.18606},
  year={2025}
}

@article{fonseca2021fsd50k,
  title={Fsd50k: an open dataset of human-labeled sound events},
  author={Fonseca, Eduardo and Favory, Xavier and Pons, Jordi and Font, Frederic and Serra, Xavier},
  journal={IEEE/ACM Transactions on Audio, Speech, and Language Processing},
  volume={30},
  pages={829--852},
  year={2021},
  publisher={IEEE}
}

\appendix

\section{Appendix: Subjective Evaluation}

\begin{figure}[h]
    \centering  \includegraphics[width=0.7\linewidth]{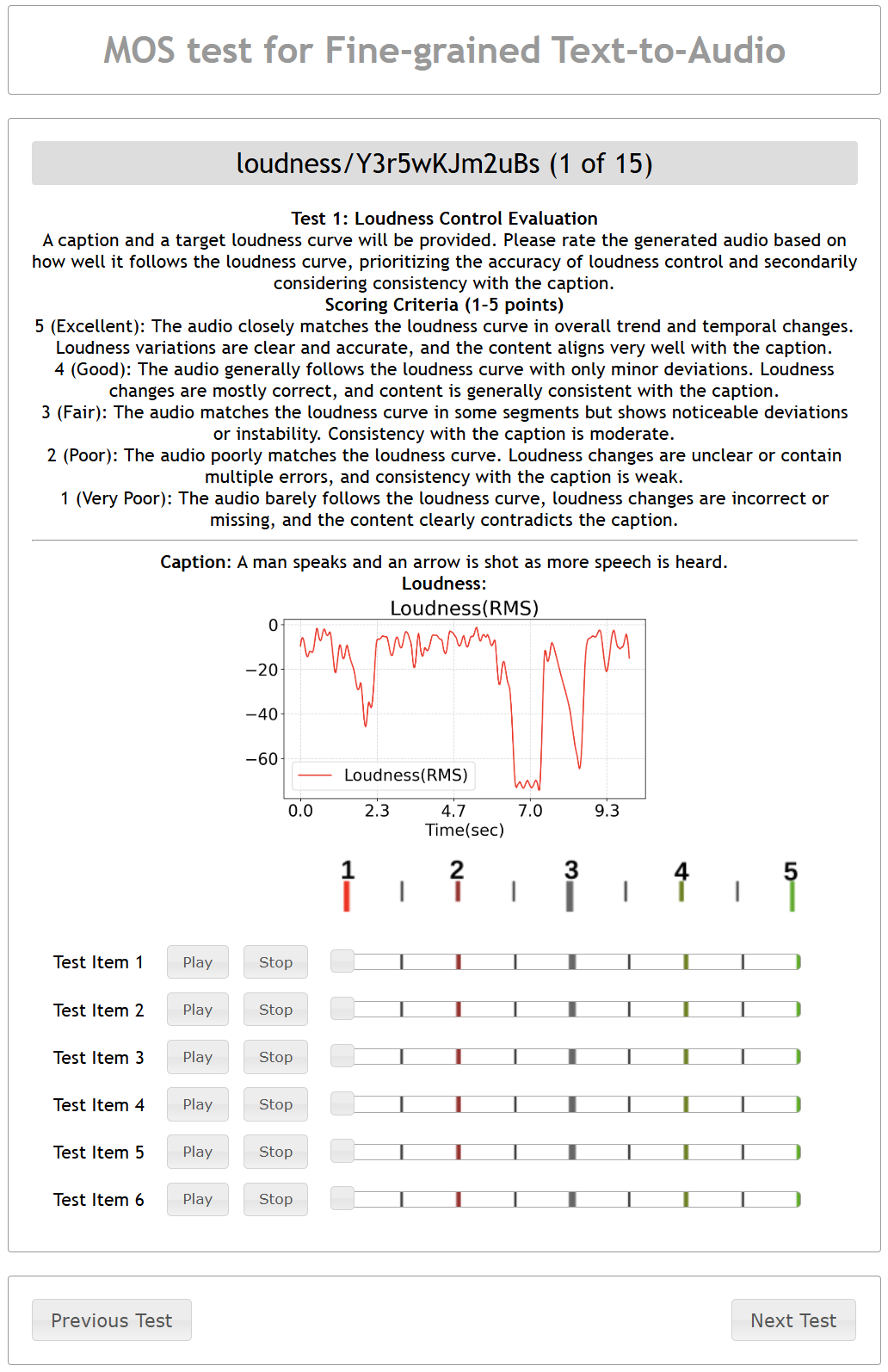}
  \caption{A snapshot of the web UI for subjective evaluation in loudness control.}
  \label{fig:mos_loudness}
\end{figure}

\begin{figure}[h]
    \centering  \includegraphics[width=0.7\linewidth]{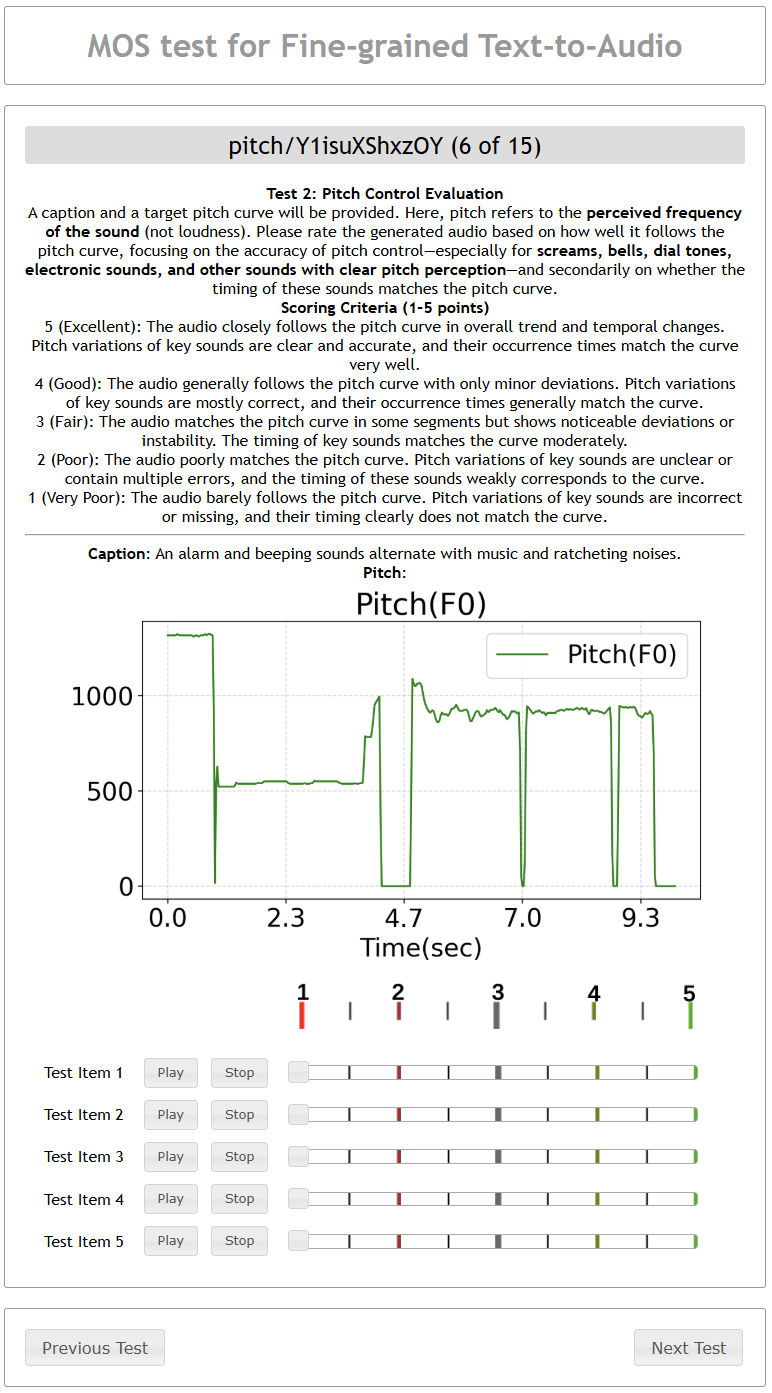}
  \caption{A snapshot of the web UI for subjective evaluation in pitch control.}
  \label{fig:mos_pitch}
\end{figure}

\begin{figure}[h]
    \centering  \includegraphics[width=0.7\linewidth]{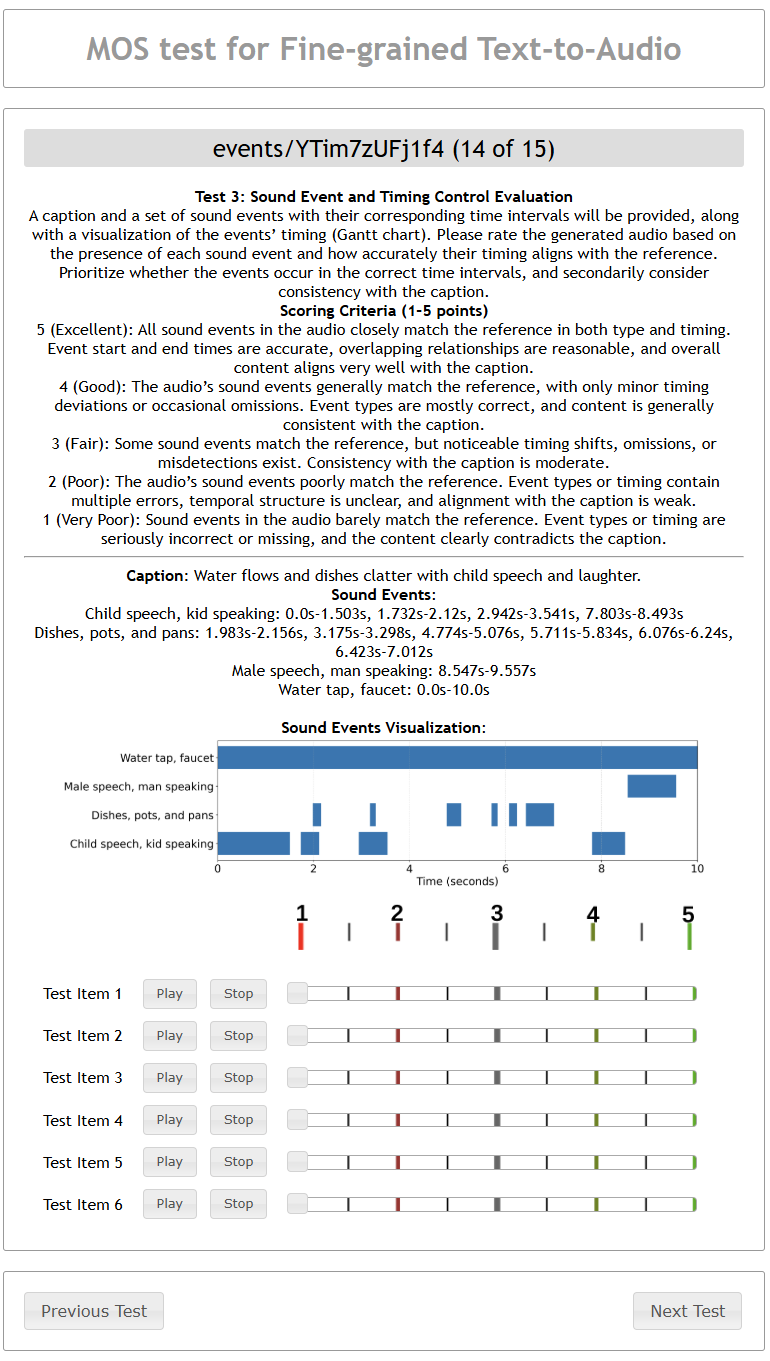}
  \caption{A snapshot of the web UI for subjective evaluation in sound events control.}
  \label{fig:mos_events}
\end{figure}

In this appendix, we provide additional details of the subjective evaluation protocol. All evaluations are conducted through a unified web-based interface, where evaluators are presented with the input caption together with the corresponding control condition (i.e., loudness curve, pitch contour, or event roll), and are asked to rate the generated audio on a 5-point MOS scale. Figures~\ref{fig:mos_loudness} \ref{fig:mos_pitch} \ref{fig:mos_events} show screenshots of the evaluation interfaces for loudness control, pitch control, and sound event control, respectively.

\section{Ethical Considerations}

Our work focuses on controllable audio generation and editing. While the models are trained on publicly available datasets, generated audio could potentially be misused for impersonation, disinformation, or creating disturbing content. To mitigate such risks, we recommend careful consideration of application contexts, adherence to relevant copyright and privacy regulations, and avoidance of sensitive or harmful use cases. We also emphasize transparency in reporting model performance and encourage the research community to adopt responsible usage guidelines.



\end{document}